\documentstyle[eqsecnum,preprint,aps,prd]{revtex} 
\preprint{DAMTP-R95/58} 
\date{\today}
\tighten
\begin{document}
\draft

\title{Entropy Perturbations due to Cosmic Strings}

\author{P. P. Avelino and R. R. Caldwell}
\address{University of Cambridge, D.A.M.T.P.\\
Silver Street, Cambridge CB3 9EW, U.K.}

\maketitle

\begin{abstract} 

We examine variations in the equation of state of the cosmic string
portion of the cosmological fluid which lead
to perturbations of the background matter density.  These
fluctuations in the equation of state are due to variations in the
local density of cosmic string loops and gravitational radiation.
Constructing a crude model of the distribution of entropy
perturbations, we obtain the resulting fluctuation spectrum using a
gauge-invariant formalism.  We compute the resulting 
cosmic microwave background anisotropy,  and estimate the effect
of these perturbations on the cosmic string structure formation 
scenario. 

\end{abstract}
\vskip 0.3in

\section{Introduction}

Cosmic strings are topological defects which may have formed in a
Grand Unified Theory era phase transition. Cosmic strings may play an
important role in cosmology \cite{COSMOstrings}, inducing temperature
anisotropies in the cosmic microwave background (CMB), and seeding
perturbations for the formation of large scale structure (LSS).

In past work, Veeraraghavan and Stebbins \cite{VS} developed a
formalism for the analysis of linearized perturbations to the
cosmological fluid induced by cosmic strings. These techniques have
been applied to the simulation of CMB temperature anisotropy \cite{CMBstrings}, 
and to the analysis of string perturbations in cold
dark matter for the formation of LSS \cite{CDMstrings}. In both
applications, the effects of the strings on the linear density field
were computed by convolving the string stress-energy tensor with the
appropriate Green's function. As well, density perturbations,
anti-correlated with the cosmic strings, were included to compensate
for perturbations induced by strings at formation.

We propose to compute the effects of an additional source of cosmic
string-generated perturbations, not explicitly included in previous
analyses. These are entropy perturbations, caused by the
inhomogeneous shifting of energy between cosmic string loops and
gravitational radiation emitted by loops. The spatial variation in
the equation of state of the total cosmological fluid is due to
statistical fluctuations in the density of long cosmic strings, from
which the distribution of small loops and gravitational radiation
follow. Below a coherence scale $\xi$, the equation of state
of the loops and radiation is uniform. The loops and radiation act
together as a dilute gas, shifting the expansion rate within a volume
$\xi^3$.  Above $\xi$, the equation of state varies as rms
statistical fluctuations, oscillating with wavelengths
longer than the coherence scale.

The reason why these effects have not been explicitly treated in past
work are the following. First, no simulation has included the
gravitational radiation emitted by loops and strings in the cosmic
string stress-energy tensor, for the purposes of CMB or LSS
calculations. The effect of gravitational wave, tensor perturbations
on the scalar, gravitational potential is a second order effect.
However, variations in the equation of state, as considered in this
paper, result in a first order effect in linearized gravity.
Second, the perturbations resulting from the
variations in the equation of state, although not genuine
fluctuations in the total density, have not been modeled by the
compensations employed to date \cite{VS}. The compensations
included in CMB and LSS analyses have simply accounted for the
initial distribution of strings and loops, not the subsequent shift
in the form of matter, from loops and long strings
to gravitational radiation.

Our goal in this paper is to estimate the importance of entropy
perturbations for the formation of large scale structures and CMB
anisotropy, prior to carrying out a more detailed investigation (some
preliminary work has been carried out in \cite{AVELINO}).  In the
following, we present a simple analysis of the effects of these
perturbations from cosmic strings.  In section \ref{model} we present a
model of the variations in the cosmic string equation of state. In
section \ref{evolution} we give explicit solutions for the
gauge-invariant potential, characterizing the density fluctuations
induced by the entropy perturbations. We also compute the power
spectrum, and comment on the effects on large scale structure
formation.  In section \ref{cmb} we estimate the effect of these
perturbations on the CMB. We conclude in section \ref{conclude}.

\section{Model of Entropy Perturbations}
\label{model}

Entropy perturbations to the cosmological fluid arise when
there is an inhomogeneous shift in energy from one 
form of matter to another. Consider a perturbing component
of the fluid $\delta \rho$ with pressure $\delta p$. The resulting
perturbation to the entropy density is
\begin{equation}
\delta S = T^{-1}(\delta p - c_s^2 \delta\rho) 
\label{makeentropy}
\end{equation}
where $c_s$ is the speed of sound in the background medium
at temperature $T$.
Hence, the presence of an additional form of matter for
which $\delta p \neq c_s^2 \delta \rho$, that is the perturbing
matter does not obey the equation of state of the background
fluid, leads to a perturbation of the entropy density.
In the cosmic string scenario, the long cosmic strings,
loops, and emitted gravitational radiation,
\begin{equation}
\delta\rho = \rho_\infty + \rho_{loops} + \rho_{gr} 
\label{totalenergy}
\end{equation}
serve as the perturbing component of the cosmological fluid,
driving the entropy perturbations in (\ref{makeentropy}).
More formally, we write
\begin{equation}
T \delta S = -\Theta_{\mu\nu} g^{\mu\nu} 
- (1 +  c_s^2) \Theta_{\mu\nu} u^{\mu} u^{\nu}  
\label{formalentropy}
\end{equation}
where $\Theta_{\mu \nu}(\eta, \vec x)$ is the cosmic string stress-energy
tensor, including the gravitational radiation,
and $u_\mu$ is a unit time-like 4-vector orthogonal
to the time-constant spatial hypersurfaces. Given the
source (\ref{formalentropy}), we may carry out a procedure
similar to that in \cite{VS,CMBstrings,CDMstrings}: convolve the source
with the appropriate Green's function to obtain the gravitational
potential. While it may be straightforward to adapt a 
cosmic string simulation to this task, in the present 
work we will construct an analytic model describing the
cosmic string entropy perturbations. 

Treating the cosmic string network as a dilute gas, we may compute the
pressure contributed by the strings. In the dust-dominated era, only
the gravitational radiation contributes to the entropy perturbations,
$\delta S$ in equation (\ref{makeentropy}).  Loops evolve as
collisionless dust, so do not contribute. Hence, it is the
shifting of energy from cosmic strings into gravitational radiation that
drives the entropy perturbations.

Based on a crude model of network evolution in which the strings
rapidly approach a scaling solution, the energy injected into
gravitational radiation, in the dust-dominated era, in a time interval
given by the coherence time scale $\xi$ is a constant fraction, 
${\cal F}$, of the background energy density. Because the gravitational
radiation is free-streamed at the speed of light, we estimate that the
coherence scale is given by the Hubble length, $\xi = {\cal H}^{-1}$.
Hence, decomposing the perturbing energy density field
in terms of an amplitude and spatial distribution, we write 
\begin{equation}
T \delta S(\eta,\vec x) = 
{1 \over 3}{\cal F}_{gr}\rho_o f(\eta,\vec x).
\label{rhosource}
\end{equation}
Here, $\eta$ is the conformal time,
$\rho_o$ is the background fluid energy density, ${\cal F}$
is the rms amplitude of the statistical fluctuations in the energy density
as a fraction of the background, and
$f(\eta,\vec x)$ is a normalized distribution describing the
spatial variation of the entropy perturbations.
In the dust-dominated era, 
${\cal F}_{gr} \sim 4\pi \widetilde A G\mu (1 - 2\langle v^2 \rangle) 
\sim O(1) G\mu$.
In this expression $\widetilde A$, defined by 
\begin{equation}
\widetilde A  \equiv \sqrt{ \langle 
( \rho_\infty - \langle \rho_\infty \rangle)_{\xi}^2 
\rangle} t^2 \mu^{-1},
\label{widetildedef}
\end{equation}
is the amplitude of the statistical fluctuations in the
long string density $\rho_\infty$, averaged on a length
scale $\xi$, where $t$ is physical time. The squared, average
velocity along the string is $\langle v^2 \rangle = 0.37$ in the
dust-dominated era \cite{numsims}.
We caution that the decomposition (\ref{rhosource})
is a simplification. A more realistic analysis would 
use the stress-energy tensor of the cosmic strings, as
determined by a numerical simulation, to compute $\delta S$. 

The function $f(\eta,\vec x)$ describes how the 
perturbations to the equation
of state vary with position and time. We decompose
this function in terms of fourier harmonics
\begin{eqnarray}
f(\eta,\vec x) &=& (2 \pi)^{-3}\int d^3 k \, 
{\rm e}^{i\vec k \cdot \vec x} \widetilde f(\eta,k) 
\cr\cr
\langle \widetilde f(\eta,k) \widetilde f(\eta',k') \rangle 
&=& (2 \pi)^3  \delta(\vec k - \vec k') 
|\widetilde f(\eta,k) \widetilde f(\eta',k)|,
\end{eqnarray}
where $\widetilde f(\eta,k) = 
|\widetilde f(\eta,k)|{\rm e}^{i \theta_{\vec k}}$.
The fluctuations in the density of long
strings lead to statistically homogeneous, isotropic variations in the
density of loops and gravitational radiation. 
Because the long cosmic strings make a statistically
random walk on scales larger than the
string correlation length, the variations in the
equation of state, and hence the phases $\theta_{\vec k}$,
are randomly distributed. The product of the fourier
coefficients, $|\widetilde f(\eta,k) f(\eta',k)|$, plays a role similar to
that of the structure function used by Albrecht \& Stebbins
\cite{CDMstrings} to describe the matter perturbations directly
induced by cosmic strings and loops.
We normalize this distribution
by smoothing the variance over a time scale $\xi$:
\begin{equation}
\sigma_f^2 \equiv \xi^{-1} \int_0^\xi d\xi' 
\langle f(\eta,\vec x) f(\eta',\vec x) \rangle = 1.
\label{normalize}
\end{equation}
The reason for smoothing in time is that the cosmological
fluid can only react to changes in the local equation of state
due to the cosmic string loops and gravitational radiation
on time scales greater than the coherence time scale. Hence
we need only restrict the variance of the distribution $f$
on the time scales which will lead to entropy perturbations.

We expect that the distribution does not vary on length scales
smaller than the time $\xi$, so that $\widetilde f(\eta, k > \xi^{-1}) = 0$. 
As a general distribution we write $\widetilde f$ as a power law in
wavenumber $k$, on scales above $\xi$:
\begin{equation} 
|\widetilde f(\eta,k)| = 2 \pi \sqrt{(\alpha +
{5 / 2})(\alpha + 3/2)} \xi^{3/2} (k \xi)^\alpha \Theta(1 - k \xi)
\end{equation} 
where $\xi/\eta =$ constant. The numerical coefficients and dependence
on $\xi$ follow as a result of the normalization, in
equation (\ref{normalize}).
The step function is defined as
$\Theta(x)=1$ for $x \ge 0$ and $\Theta(x)=0$ for $x = 0$.
The parameter $\alpha$ determines how scales above the 
coherence scale $\xi$ contribute to the distribution.
As $\alpha \to \infty$ 
\begin{equation}
\lim_{\alpha \to \infty} |\widetilde f(\eta, k)| = 2 \pi 
\xi^{3/2} \delta(1 - k \xi) 
\end{equation} 
so that the distribution is characterized by a single, comoving scale.
For finite, decreasing values of $\alpha$, the range of comoving scales
contributing to $\widetilde f$ increases. In the limit $\alpha\to 0$,
the distribution has support on all wavelengths above the comoving
scale $\xi$.  Because the network of long strings is distributed as a
random walk on large scales, we expect the entropy perturbations to
similarly have support on scales above $\xi$.
Hence, we expect $\alpha \to 0$ to be more physically reasonable than
$\alpha \to \infty$.  However, in the following calculations we will
take $\alpha \to \infty$ in order to simplify the calculations, noting
that for other values of $\alpha$, the results are not strongly
affected.  Having specified the properties of the entropy
perturbations, we may proceed to determine the resulting fluctuation
spectrum.

\section{Evolution of Perturbations}
\label{evolution}

To describe the perturbations of the background cosmological fluid, we
use a gauge-invariant formalism \cite{MFB,DURRER}.  We must supply an
expression for the density perturbation relative to the hypersurface
which represents the local rest frame of the cosmological fluid; this
prescription gives as close as possible a ``Newtonian'' time slicing.
For an entropy perturbation, that is a perturbation to the equation of
state, we find the evolution of the gauge-invariant potential $\Phi$ is
given by
\begin{equation}
\Phi'' + 3 (1 + c^2_s) {\cal H} \Phi' + 
[2{\cal H'} + (1 + 3 c^2_s) {\cal H}^2] \Phi- c^2_s \nabla^2\Phi
= 4 \pi G a^2 (\delta p - c^2_s \delta \rho). 
\label{waveeqn}
\end{equation}
Here we work with conformal time $\eta$, where $'\equiv
\partial/\partial\eta$, and ${\cal H} = a'/a$.  The coordinates
$(\eta,\vec x)$ are dimensionless, as the expansion scale factor $a$
carries units of length.

We will also compute the power spectrum of the
gauge-invariant potential generated by the entropy
perturbations in the dust-dominated era.
We define the correlation function of the potential to be
the second moment of $\Phi$:
\begin{equation}
\xi(\eta,r)  \equiv  \langle \Phi(\eta,\vec x)\Phi(\eta,\vec x') \rangle 
= 4 \pi \int k^2 dk \, {\sin(k r) \over k r} {\cal P}_{\Phi}(\eta,k) 
\label{power}
\end{equation}
Here $r = |\vec x - \vec x'|$ is a conformal distance.
Recall that the density contrast $\delta$ due to the 
gauge-invariant potential is given by
\begin{equation}
\delta \equiv {2 \over 3} {\cal H}^{-2} \Big( \nabla^2 \Phi  - 3 {\cal H}\Phi'
 - 3 {\cal H}^2 \Phi \Big).
\end{equation}
For modes well within the horizon, $k \gg {\cal H}$, the density contrast
is dominated by the gradient term. Thus, we have the relationship
\begin{equation}
{\cal P}_\delta = {4 k^4 \over 9 {\cal H}^4} {\cal P}_\Phi
\qquad {\rm for} \quad k \gg {\cal H}
\label{densitypower}
\end{equation}
relating the power spectrum in the 
potential to the power in the density contrast.

In the dust-dominated era, the scale factor behaves as $a \propto \eta^2$
and the speed of sound $c_s$ for the cold, collisionless
dust vanishes so that the evolution
equation for the gauge-invariant potential simplifies
greatly. We may solve the differential equation as
\begin{eqnarray}
\Phi(\eta,\vec x) =  4 \pi G &&
\int_{\eta_{eq}}^\eta d\eta' \, \eta'^{-6} 
\int_{\eta_{eq}}^{\eta'} d\eta'' \, \eta''^{6} a^2(\eta'') 
\Big( \delta p(\eta'',\vec x) - c_s^2 \delta \rho(\eta'',\vec x) \Bigr)
\cr\cr
&& + (\eta - \eta_{eq})\Phi'(\eta_{eq},\vec x) + \Phi(\eta_{eq},\vec x).
\label{dustphi}
\end{eqnarray}
Hence, given the boundary terms $\Phi$ and $\Phi'$ at $\eta_{eq}$,
at the onset of the dust-dominated era, it is straight forward
to obtain the potential $\Phi$ at any later time.

The source of the entropy perturbations is the gravitational radiation
emitted by oscillating cosmic string loops.  Using (\ref{rhosource}),
we may obtain the dust-era solution for $\Phi$.  We will make the
simplification that the boundary terms are unimportant. This is
reasonable provided we are interested in length scales due to
perturbations which enter the horizon after radiation-matter equality.
Hence, we find
\begin{equation}
{\cal P}_\Phi(\eta,k) = {4 {\cal F}_{gr}^2 \over (2 \pi)^3}
\Big| \int_{\eta_{eq}}^{\eta} d\eta' \, (\eta')^{-6}
\int_{\eta_{eq}}^{\eta'} d\eta'' \, (\eta'')^{4} \widetilde f(\eta'',k)
\Big|^2.
\end{equation}
In the case $\alpha\to\infty$, the power is
\begin{equation}
\lim_{\alpha \to \infty}
{\cal P}_\Phi(\eta,k) =  
{2 {\cal F}_{gr}^2 \over 25 \pi}
k^{-3} \big[ 1 - x^{-5}  \big]^2
\Theta(x-1)\Theta(1-x_{eq}).
\label{deltadust}
\end{equation}
Here, $x = k \xi$.  For other values of $\alpha$ the power spectrum
also behaves as ${\cal P}_\Phi \propto k^{-3}$ for modes well within
the horizon, generating a scale-invariant isocurvature spectrum.
Fluctuations generated during the dust-dominated era will have
wavelengths from $\sim 10 h^{-2}$Mpc, the horizon scale at
radiation-matter equality, up to the present horizon scale.  These
fluctuations will contribute to the large angle CMB anisotropy and the
formation of the largest structures.

We may compare the power in entropy-generated density fluctuations with
the power due to the perturbations induced directly by the cosmic
strings. Referring to (\ref{densitypower}), we find
\begin{equation}
4 \pi k^3 {\cal P}_{\delta}(k) = {288 \over 25} {\cal F}_{gr}^2 (k h)^4
\end{equation}
where $k$ has units $h^2$/Mpc and $h$ is the Hubble parameter
normalized to $H_0 = 100$km/s/Mpc. Compare this to the power spectrum
due to cosmic strings in CDM (equation 8 of \cite{CDMstrings}):
\begin{equation}
4 \pi k^3 {\cal P}_{\delta}(k) = 4 \pi (6.8)^2 (k h)^4 (G\mu)^2.
\end{equation}
Here we have used the I-model  of Albrecht \& Stebbins
\cite{CDMstrings} motivated by the string simulation results of Bennett
\& Bouchet and Allen \& Shellard \cite{numsims}, on length scales well
above the $8 h^{-1}$Mpc normalization scale. Constructing a ratio of
the LSS power due to the perturbations induced directly in the
cosmological fluid by cosmic strings, to those generated by entropy
perturbations, we obtain
\begin{equation}
{\cal P}_{direct} / {\cal P}_{entropy} 
\sim 50 (G\mu)^2 {\cal F}_{gr}^{-2} = 3.2 \widetilde A^{-2}.
\label{ratio}
\end{equation}
We see that for ${\cal F}_{gr} \sim 7 G\mu$ or, using the estimate for
${\cal F}_{gr}$ preceding equation (\ref{widetildedef}), for
$\widetilde A \sim 1.8$ the power for large scale structures are
comparable.  As $\alpha \to 0$, such that the entropy perturbations
have support on a wide range of length scales, the power due to entropy
perturbations increases slightly, and the powers are comparable for
$\widetilde A \sim 1.4$. The reason for the increase is that entropy
perturbations generated at a range of times, rather than at a single
time as with the $\alpha\to\infty$ model, now contribute to the power
in a single mode, on these scales.  Numerical simulations indicate that
the mean energy density in long strings during the dust-dominated era
is $\rho_\infty = (4 \pm 1)\mu/t^2$ \cite{numsims}, obtained by
averaging over multiple realizations of the string simulation.  If the
quoted uncertainty in $\rho_{\infty}$ measures the amplitude of the
statistical fluctuations, averaged on a scale $\xi$, then $\widetilde A
\sim 1$.  Modeling the long string network as a random walk on large
scales, we expect $\widetilde A \sim 0.6$ \cite{AVELINO}.  Hence, we
expect $\widetilde A \sim 0.6 - 1$ to be a reasonable range for the
amplitude of entropy perturbations.  We caution that these results are
intended as a preliminary estimate; a more reliable calculation will
use the string simulations directly to compute the effects of entropy
perturbations.

\section{CMB Anisotropy}
\label{cmb}

The CMB anisotropies induced by the entropy-generated perturbations may
be computed from the gauge-invariant potential.  The dominant
contribution on large angular scales, as is common for isocurvature
perturbations, is due to the time variation of
the potential $\Phi$, integrated along the line of sight to the surface
of last scattering:
\begin{equation}
{{\Delta} T \over T}(\hat n) =  
2 \int_{\eta_{ls}}^{\eta_o} d\eta \, \Phi'(\eta, |\eta_o - \eta|\hat n).
\end{equation}
Here, $\hat n$ is a unit-vector on the celestial sphere.  
The correlation function is expressed in the usual form,
\begin{equation}
C(\theta) \equiv \langle {{\Delta} T \over T}(\hat n)
{{\Delta} T \over T}(\hat n') \rangle = 
\sum_{\ell=2}^\infty {2\ell + 1 \over 4 \pi} C_\ell P_\ell(\cos\theta).
\end{equation}

The amplitude of the multipole moments is given by $C_\ell$, and $\hat
n \cdot \hat n' = \cos\theta$.  Evaluating the large-angle CMB
anisotropy due to dust-era entropy perturbations only, in the
$\alpha\to\infty$ case, we find $\ell^2 C_\ell \sim 6 {\cal F}_{gr}^2$
for $10 \lesssim \ell \lesssim 40$.  Recall that ${\cal F}_{gr} \sim 4
\widetilde A G\mu$, so that for $\widetilde A \sim 2.0$ these multipole
moments are comparable to the $\ell^2 C_\ell \sim 400 (G\mu)^2$
perturbations measured in the numerical simulations \cite{CMBstrings}.
This is roughly consistent with the result for $\widetilde A (\sim
1.8)$ found in the previous section, in equation (\ref{ratio}), in
order that the LSS power due to the entropy perturbations is comparable
to the power due to the perturbations induced directly by cosmic
strings.

The effects of the entropy-generated perturbations on small-scale CMB
anisotropy are estimated using the analytic techniques developed by Hu
\& Sugiyama \cite{HUSUGIYAMA}. 
\begin{eqnarray}
C_\ell &=& {2 \over \pi} \int {dk \over k}
\, k^3 | \theta_\ell(\eta_o,k)|^2 \cr\cr
\theta_\ell(\eta_o,k) &=&
\Big( \theta_0(\eta_{ls},k) - \widetilde \Phi(\eta_{ls},k) \Big)
j_\ell(k|\eta_o - \eta_{ls}|) \cr\cr
&&+ \theta_1(\eta_{ls},k) (2\ell +1)^{-1}
\Big( \ell j_{\ell-1}(k|\eta_o - \eta_{ls}|) - 
(\ell+1) j_{\ell+1}(k|\eta_o - \eta_{ls}|) \Big) \cr\cr
&&- 2 \int_{\eta_{ls}}^{\eta_o} d\eta \widetilde \Phi'(\eta,k)
j_\ell(k|\eta_o - \eta|)
\end{eqnarray}
In the above equations (see equations 12 and 14 of \cite{HUSUGIYAMA}),
$\theta_0$ and $\theta_1$ are the monopole and dipole moments of the
perturbations to the baryon-photon fluid. The final term, the
integrated Sachs-Wolfe effect, dominates on large angular scales, as
with isocurvature models. Modeling only the dust-era entropy
perturbations, the resulting contribution to the CMB anisotropy due to
monopole and dipole oscillations of the baryon-photon fluid is small.
However, if we artificially lengthen the duration of the
pre-recombination, matter-dominated epoch as a way of extending the
validity of our model to smaller scales and earlier times, we find that
significant anisotropy may result on small angular scales.  For the
$\alpha \to \infty$ model with a coherence scale $\xi = {\cal H}^{-1}$,
a series of peaks develops, beginning near $\ell \gtrsim 200$, before
diffusion damps out the perturbations. Smaller values of the coherence
scale lead to a decrease in the the peak amplitude and a shift in the
peak location to higher $\ell$ values. It is important to note that the
location of the peak in the spectrum is sensitive to the value of the
comoving coherence scale $\xi$ for the entropy perturbations. Magueijo
{\it et al} \cite{MAGUEIJO} have carried out a similar calculation of the
small-angle anisotropies using the Albrecht-Stebbins LSS model as an
effective source for fluctuations in the baryon-photon fluid. There,
they found that the location of the peak shifts to larger values of
$\ell$ as the coherence length scale decreases.  This behavior may be
an important tool for discriminating between inflationary and
defect-based models.  Furthermore, perturbations generated by
scale-dependent variations in the equation of state, as considered in
this paper, may be a generic property of defect-based cosmological
scenarios.  For global defects, however, the generation of entropy
perturbations will be due to the distribution of Goldstone, rather than
gravitational, radiation.  Similar calculations have been carried out
for textures by Crittenden \& Turok \cite{CRITTENDEN}.

\section{Conclusion}
\label{conclude}

In this paper we have analyzed the effects of entropy perturbations in
the cosmic string scenario. These perturbations arise from spatial
variations in the density of cosmic string loops and gravitational
radiation, leading to a field of approximately Gaussian fluctuations in
the equation of state of the cosmological fluid. This source of
anisotropy has not been explicitly considered in previous analyses,
because the contribution to the cosmic string stress-energy tensor by
gravitational waves has not been included. By constructing a crude
model of this source, we have been able to estimate the effects of the
resulting perturbations on large scale structure formation and the
cosmic microwave background.

Let us comment now on this analytic model. We claim that the
distribution function used to describe the variations in the density of
gravitational radiation and loops, and hence the variations in the
equation of state, is a reasonable first approximation for the purposes
of estimating the relative importance of this effect. There are certain
shortcomings of this construction, among them the uncertainties in the
amplitude of the rms fluctuations $\widetilde A$, the size of the
comoving coherence scale $\xi$, and the dependence of the distribution
$f$ on wavenumber $k$, which we have condensed into the parameter
$\alpha$. For large scale perturbations, the amplitude and shape of the
spectrum does not depend strongly on $\alpha$ or $\xi$, as a
scale-invariant isocurvature power spectrum results. For what we expect
to be a reasonable range for the amplitude, $\widetilde A \sim 0.6 -
1$,  the entropy perturbations make a  non-negligible contribution  to
the LSS and large-angle CMB spectrum.  In summary, we have crudely
modeled the essential features; a statistically homogeneous, isotropic
distribution of fluctuations in the equation of state, varying on
length scales above the comoving horizon radius. It remains to compare
this model with the distribution observed in a realistic cosmic string
simulation.  We plan to carry out these advances in the near future.

\acknowledgements
We thank Paul Shellard for useful conversations.
The work of PPA was supported by JNICT.
The work of RRC was supported by PPARC through grant number GR/H71550.


\end{document}